\documentclass[english]{article}
\usepackage[T1]{fontenc}
\usepackage[utf8]{inputenc}
\usepackage{textcomp}
\usepackage{amsmath}
\usepackage{graphicx}
\usepackage{esint}

\makeatletter
\newcommand{\lyxaddress}[1]{
\par {\raggedright #1
\vspace{1.4em}
\noindent\par}
}

\@ifundefined{showcaptionsetup}{}{%
 \PassOptionsToPackage{caption=false}{subfig}}
\usepackage{subfig}
\makeatother

\usepackage{babel}

\begin{document}

\title{On the average of electrostatic and magnetostatic fields, the singularities
of dipole fields and depolarizing fields}

\author{Patrick De Visschere}

\maketitle

\lyxaddress{\noindent \begin{center}
Ghent University -- ELIS\\
Sint-Pietersnieuwstraat 41, B-9000 Gent, Belgium\\
pdv@elis.ugent.be
\par\end{center}}
\begin{abstract}
In {}``Classical Electrodynamics'' \cite{jackson} a theorem is
proved on the average of an electrostatic or magnetostatic field over
a spherical volume. The proof of the theorem is based on an expansion
in spherical harmonics and it is useful for deriving the singular
behaviour of dipole fields. In this paper we give a simple proof of
this theorem, the key element being to use reciprocity. The method
also highlights the relation between the dipole singularity and the
depolarizing field of a sphere, which is thus also found very easily.
In addition this link enables us to extend the theorem to a more general
ellipsoidal volume. Unfortunately one can only prove the existence
of the depolarizing tensor in this way and one still needs another
method for actually calculating the tensor. Finally the theory is
also extended to an anisotropic background medium.
\end{abstract}

\section{Introduction.}

In \cite{jackson} it is proved that the average of the electrostatic
field taken over a spherical volume equals the electric field at the
center of the sphere if the charge distribution causing the field
is completely located outside the sphere. If on the other hand the
charge distribution is completely enclosed by the sphere then the
average is proportional to the elecric dipole moment of the charge
distribution calculated with respect to the center of the sphere.
The proof of this property is based on an expansion in spherical harmonics
of the potential of a monopole and similar results hold for magnetostatic
fields. The second form of the theorem has then be used to derive
the singularity of the electric/magnetic dipole fields, which has
received some attention recently \cite{singularities,singularities2}.

In §\ref{sec:Electrostatic-field.}-\ref{sec:Magnetostatic-field.}
we will show that both theorems can be proved rather easily by invoking
the reciprocity of the fields and especially for the electrostatic
case (§\ref{sec:Electrostatic-field.}) the derivation is almost immediate.
In §\ref{sec:Magnetostatic-field.} the corresponding magnetostatic
theorem is proved using the conventional current loop model for a
magnetic dipole. In §\ref{sec:Singularities-of-the} we briefly recall
how the singularities of the dipole fields can be found using these
theorems. In §\ref{sec:Depolarization-fields.} the close link is
revealed between these singularities and the well-known depolarizing
field of a sphere. This also enables us to extend the results to more
general shapes of the volume and finally in §\ref{sec:Anisotropic-medium.}
to an anisotropic background medium.

\section{\label{sec:Electrostatic-field.}Electrostatic field.}

\begin{figure}
\subfloat[The charge is located outside of the sphere.]{

\includegraphics[scale=0.7]{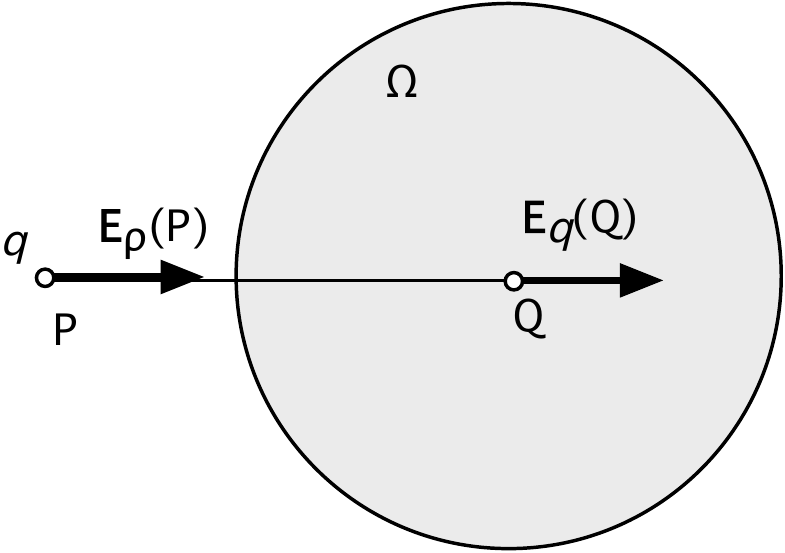}}\hfill{}\subfloat[The charge is located inside the sphere.]{

\includegraphics[scale=0.7]{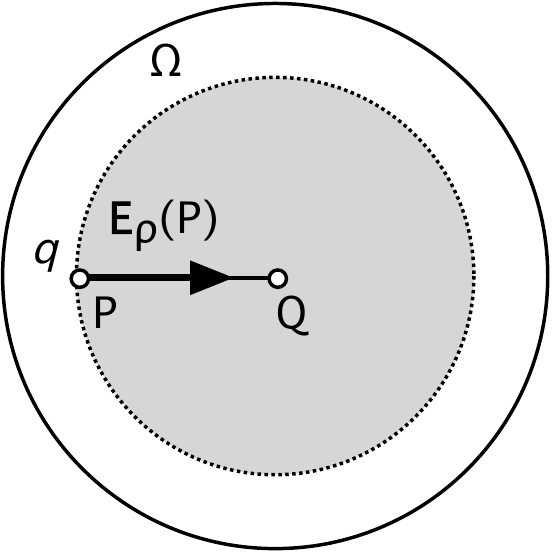}}

\caption{\label{fig:E}The average over a spherical volume $\Omega$ of the
electrostatic field of a monopole \emph{q} is equal to the electrostatic
field of a uniform charge density $\rho=-q/|\Omega|$ in the sphere
and measured at the location of the monopole.}

\end{figure}

It is of course sufficient to prove the theorem for the field of a
single monopole \emph{q}. Using superposition the extension to an
arbitrary charge distribution is then straightforward. We thus consider
the average of the field of a monopole $q$, located at P, over a
spherical volume $\Omega$ (see figure~\ref{fig:E}). Using reciprocity
we conclude that the average field $\left\langle \overline{E}_{q}\right\rangle _{\Omega}$
equals the field $\overline{E}_{\rho}(\mathrm{P})$ at the source
point P generated by a charge $-q$ uniformly distributed over the
volume $\Omega$ with density $\rho=-q/|\Omega|$, where $|\Omega|$
is the size of this volume. This first step in the reasoning is valid
for an arbitrary volume. For a spherical volume with center at Q the
latter field is easily calculated. If the source point P is outside
the sphere ($\mathrm{P}\notin\Omega$) then $\overline{E}_{\rho}(\mathrm{P})$
equals the field of a single monopole $-q$ at the center Q of the
sphere and applying reciprocity again it follows that the averaged
field equals the field of the original charge \emph{q} measured at
the center of the sphere: \begin{equation}
\left\langle \overline{E}_{q}\right\rangle _{\Omega}=\overline{E}_{\rho}(\mathrm{P})=\overline{E}_{-q}(\mathrm{P})=\overline{E}_{q}(\mathrm{Q})\qquad\mathrm{P}\notin\Omega\label{eq:1}\end{equation}
The two outer equalities are due to reciprocity, the central one is
due to the spherical symmetry.

If on the other hand the source point is enclosed by the sphere ($\mathrm{P}\in\Omega$)
then invoking again the spherical symmetry a standard application
of Gauss integral law yields: \begin{equation}
\left\langle \overline{E}_{q}\right\rangle _{\Omega}=\overline{E}_{\rho}(\mathrm{P})=-\frac{q\overline{\mathrm{QP}}}{3\varepsilon_{0}|\Omega|}=-\frac{\bar{p}_{q}}{3\varepsilon_{0}|\Omega|}\qquad\mathrm{P}\in\Omega\label{eq:2}\end{equation}
where $\overline{p}_{q}$ is the dipole moment of $q$ with respect
to the center of the sphere.

\section{\label{sec:Magnetostatic-field.}Magnetostatic field.}

A similar reasoning can be followed for the magnetostatic case, although
the calculations are slightly more complicated if the conventional
electric-current model is used for the magnetic dipole ($\overline{m}=i\bar{a}$
with \emph{i} the current in the loop and $\overline{a}$ the oriented
area of the loop). The magnetostatic result can be obtained much more
economically by using the magnetic-charge model \cite{chu} instead
(see §\ref{sec:Singularities-of-the}). However since the current
model is more widely adopted and for completeness we show the derivation
for this model too.

\begin{figure}
\subfloat[The current element is located outside the sphere.]{

\includegraphics[scale=0.7]{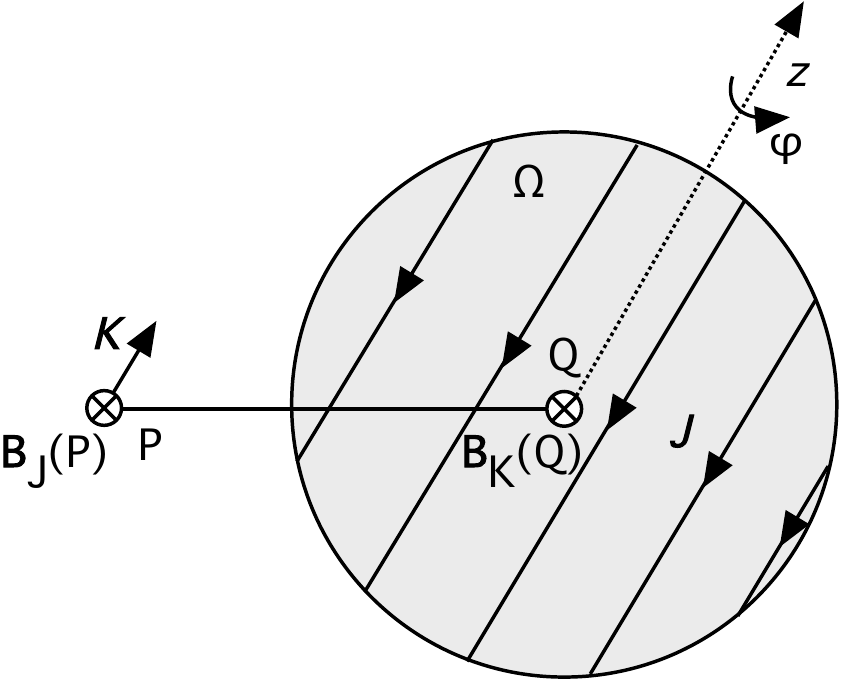}}\hfill{}\subfloat[The current element is located inside the sphere.]{

\includegraphics[scale=0.7]{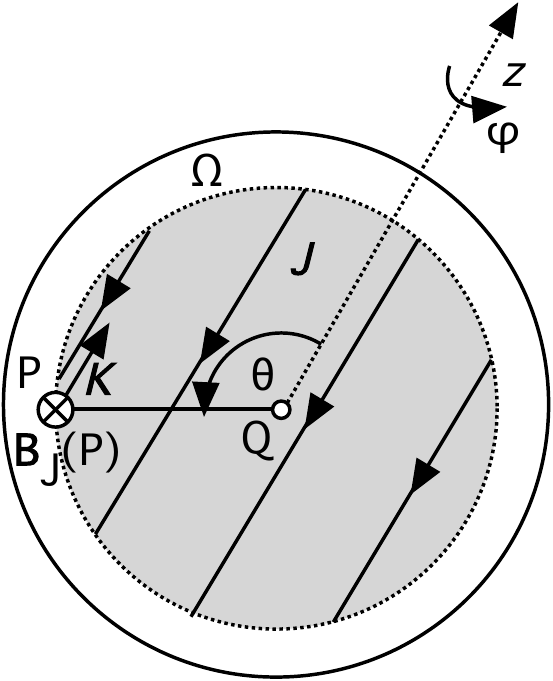}}

\caption{\label{fig:M}The average over a spherical volume $\Omega$ of the
magnetostatic field of a current element \emph{$\overline{K}$} is
equal to the magnetostatic field of a uniform current density $\bar{J}=-\overline{K}/|\Omega|$
in the sphere and measured at the location of the current element.}

\end{figure}

We consider a $\delta$-like current element which we denote by $\overline{K}$
(with the current density $\bar{J}(\bar{r})=\overline{K}\delta(\bar{r}-\bar{r}_{\mathrm{P}})$).
Without loosing generality we choose the origin at the center Q of
the sphere and the \emph{z}-axis along the direction of $\overline{K}$
with $\overline{K}=K\overline{1}_{z}$ (see figure~\ref{fig:M}).
Using the reciprocity of Biot-Savart's law the magnetic induction
$\left\langle \overline{B}_{\bar{K}}\right\rangle _{\Omega}$ of the
current element at P averaged over the sphere $\Omega$\emph{ }equals
the induction $\overline{B}_{\bar{J}}(P)$ measured at P of a uniform
current density $\overline{J}=-\overline{K}/|\Omega|$ present in
the sphere. The solution of the latter problem is analogous with the
electrostatic problem with the correspondence $q/\varepsilon_{0}\leftrightarrow\mu_{0}K$
and $E_{r}\leftrightarrow B_{\varphi}/\sin\theta$ where $E_{r}=-d\phi/dr$
is the radial electrostatic field and $B_{\varphi}=-\sin\theta dA_{z}/dr$
the tangential component of the induction, $\phi$ being the electrostatic
potential and $\overline{A}=A_{z}\overline{1}_{z}$ the vector potential,
and where $\theta$ is the angle between the $z$-axis and the position
vector $\overline{r}$. It then follows readily that:\begin{equation}
\left\langle \overline{B}_{\bar{K}}\right\rangle _{\Omega}=\overline{B}_{\bar{J}}(\mathrm{P})=\bar{B}_{-\bar{K}}(\mathrm{P})=\overline{B}_{\bar{K}}(\mathrm{Q})\quad\mathrm{P}\notin\Omega\label{eq:3}\end{equation}

for the situation in figure~\ref{fig:M}(a) and for the situation
in figure~\ref{fig:M}(b) we find similarly:\begin{equation}
\left\langle \overline{B}_{\bar{K}}\right\rangle _{\Omega}=\overline{B}_{\bar{J}}(\mathrm{P})=-\frac{\mu_{0}K}{3|\Omega|}|\overline{\mathrm{PQ}}|\sin\theta\overline{1}_{\varphi}=-\frac{\mu_{0}}{3|\Omega|}\overline{\mathrm{PQ}}\times\overline{K}=\frac{2\mu_{0}}{3|\Omega|}\overline{m}_{\bar{K}}\quad\mathrm{P}\in\Omega\label{eq:4}\end{equation}

where $\overline{m}_{\bar{K}}$ is the magnetic moment of the current
element with respect to the center Q of the sphere. Note that the
extra factor 2 compared with the electrical dipole is due to the factor
1/2 in the definition of the magnetic moment, this being a second
order moment, whereas the electric dipole moment is of first order.
The results \eqref{eq:3} and \eqref{eq:4} are again easily extended
for an arbitrary current density distribution $\overline{J}(\overline{r})$.

\section{\label{sec:Singularities-of-the}Singularities of the dipole fields.}

Applying the second form of the theorem \eqref{eq:2} to the field
of a dipole $\bar{p}$ enclosed by an arbitrary sphere it follows
that the field must contain a singularity $-\frac{\bar{p}}{3\varepsilon_{0}}\delta(\bar{r})$.
To take this singularity into account the field is then written explicitly
as \cite{jackson}:\begin{equation}
\varepsilon_{0}\overline{E}_{\overline{p}}=\frac{\overline{p}}{4\pi}\cdot\frac{3\overline{r}\overline{r}-r^{2}\overline{\overline{I}}}{r^{5}}-\frac{\overline{p}}{3}\delta(\overline{r})\label{eq:dipole}\end{equation}

where $\overline{\overline{I}}$ is the unit tensor and where we have
written the field in a form which is more amenable to a generalization
to anisotropic media (see §\ref{sec:Anisotropic-medium.}). The interpretation
of this expression requires some care. If the sphere is centered on
the dipole then the average of the first non-singular term vanishes
only if one excludes an infinitesimal spherical volume around the
dipole. This actually means that \eqref{eq:dipole} is only valid
if the dipole is modelled as an infinitesimally small polarized sphere,
as will become more explicit in §\ref{sec:Depolarization-fields.}.
As an additional bonus we find that, since the average of the field
is, according to \eqref{eq:2}, independent of the position of the
dipole within the sphere, the principal value integral of the first
term must also vanish if the sphere is not centered on the dipole.

The corresponding magnetostatic result can be obtained immediately
by using the magnetic-charge model \cite{chu} for the magnetic dipole.
In this model a magnetic moment is composed of 2 magnetic charges
$\pm\, q^{*}$ a distance $\overline{d}$ apart. Then $\mu_{0}\overline{m}=q^{*}\overline{d}$
and due to the analogy with the electrostatic case the magnetic field
$\overline{H}$ will show a singularity $-\frac{\overline{m}}{3}\delta(\bar{r})$.
Since $\overline{B}/\mu_{0}=\overline{H}+\overline{M}$ and $\overline{M}=\overline{m}\delta(\bar{r})$
it follows that $\overline{B}/\mu_{0}$ must have a singularity $\frac{2}{3}\bar{m}\delta(\bar{r})$
which agrees with \eqref{eq:4} found higher. Finally we note that
for the electrostatic case the electric induction $\bar{D}$ also
shows a singularity $\frac{2}{3}\overline{p}\delta(\overline{r})$.

\section{\label{sec:Depolarization-fields.}Depolarizing fields.}

Applying the theorem in \eqref{eq:1} once more to a dipole%
\footnote{In what follows we limit ourselves to the electrostatic case, but
all results are readily extended to the magnetic case.%
} $\bar{p}$ we find that the field of a dipole averaged over a sphere
(not enclosing the dipole) is equal to the field of the same dipole
measured at the center of the sphere. Applying reciprocity to both
sides of this equality it follows that the (external) field of a uniformly
polarized sphere ($\overline{P}=\overline{p}/|\Omega|$) is equal
to the field of a dipole $\bar{p}$ at its center. The same reasoning
can be applied for any higher order multipole. Similarly applying
\eqref{eq:2} for a dipole but now enclosed by the sphere it follows
that the electric field within a uniformly polarized sphere ($\overline{P}=\overline{p}/|\Omega|$)
is uniform and equal to $-\frac{\overline{P}}{3\varepsilon_{0}}$.
This is the well-known depolarizing field of a sphere which is thus
shown to be directly related to the singularity in \eqref{eq:dipole}
and this leads us to the conclusion that \eqref{eq:dipole} is only
valid if the dipole is modelled as an infinitesimally small polarized
sphere. Applying the same reasoning to any higher-order multipole
and since the dipole moment of any higher-order multipole vanishes
we conclude that in a sphere filled with a uniform higher-order multipole
density the internal field vanishes.

These results can be formalized as follows. Using the same notation
as before the field of a multipole averaged over an arbitrary volume
$\Omega$ can be found by filling $\Omega$ with a corresponding multipole
density and looking for the field at the source point P:\begin{alignat}{1}
\left\langle \overline{E}_{q}\right\rangle _{\Omega} & =\overline{E}_{\rho}(\mathrm{P})\nonumber \\
\left\langle \overline{E}_{\overline{p}}\right\rangle _{\Omega} & =\overline{E}_{\overline{P}}(\mathrm{P})\label{eq:reciprocity}\\
\left\langle \overline{E}_{\overline{\overline{q}}}\right\rangle _{\Omega} & =\overline{E}_{\overline{\overline{Q}}}(\mathrm{P})\nonumber \\
 & \ldots\nonumber \end{alignat}

where $\rho=-q/|\Omega|$, $\overline{P}=\overline{p}/|\Omega|$,
$\overline{\overline{Q}}=-\overline{\overline{q}}/|\Omega|$, ...
are the charge density, polarization density and quadrupole density,
... corresponding with the charge $q$, the dipole $\overline{p}$,
the quadrupole $\overline{\overline{q}}$ ... . The multipole fields
on the left sides follow from each other by differentiation:\begin{alignat}{1}
\overline{E}_{\overline{p}} & =-\overline{p}\cdot\nabla\overline{E}_{q}/q\nonumber \\
\overline{E}_{\overline{\overline{q}}} & =\overline{\overline{q}}\cdot\nabla\nabla\overline{E}_{q}/q\label{eq:multipolerelations}\\
 & \ldots\nonumber \end{alignat}
Due to reciprocity similar relations hold for the fields on the right
sides:\begin{alignat}{1}
\overline{E}_{\overline{P}} & =-\overline{P}\cdot\nabla\overline{E}_{\rho}/\rho\nonumber \\
\overline{E}_{\overline{\overline{Q}}} & =\overline{\overline{Q}}\cdot\nabla\nabla\overline{E}_{\rho}/\rho\label{eq:seven}\\
 & \ldots\nonumber \end{alignat}

with $\nabla$ operating on the position vector of the (original)
source point $\mathrm{P}$. Note that \eqref{eq:reciprocity} and
\eqref{eq:seven} are valid for arbitrary volumes. The field of a
uniformly polarized body in particular can thus be derived from the
field of the same uniformly charged body. This method for calculating
the field of a uniformly polarized (or magnetized) volume has been
attributed to Poisson \cite{osborn}.

For a sphere in addition the symmetry can be invoked and the external
field of a uniformly charged sphere simply equals that of a monopole.
Therefore the field of a uniformly polarized sphere also equals a
dipole field and similar results hold for higher order multipole moments.
Since the internal field of a uniformly charged sphere is linear in
the position, as shown in \eqref{eq:2}, the internal field of a uniformly
polarized sphere is then according to the first equation in \eqref{eq:seven},
a constant and given by:\begin{equation}
\left\langle \overline{E}_{\overline{p}}\right\rangle _{\Omega}=-\frac{\bar{p}}{3\varepsilon_{0}|\Omega|}=-\frac{\overline{P}}{3\varepsilon_{0}}=\overline{E}_{\overline{P}}(\mathrm{P})\qquad\mathrm{P}\in\Omega\label{eq:depolsphere}\end{equation}

where we stress once again the relation between the singularity of
the dipole field on the left and the depolarizing field of a uniformly
polarized sphere on the right. Due to the linear dependence on the
position vector it follows immediately that the internal field vanishes
for all higher-order uniform multipole densities.

Having found this relation it becomes straightforward to look for
extensions of the theorems \eqref{eq:1} and \eqref{eq:2} to other
than spherical volumes. It is obvious that an extension to ellipsoidal
volumes must be possible for an internal point. Therefore we consider
firstly a (centric) hollow sphere. From the linear position dependence
in \eqref{eq:2} it is clear that the averaged electric field of any
charge enclosed by the hollow sphere is zero and this is equivalent
with the well-known property that the field of a uniformly charged
hollow sphere is zero in the enclosed space. It also readily follows
that this remains true for an enclosed dipole distribution (and for
the corresponding polarized hollow sphere) even if the inner sphere
is eccentric. This argument can be reversed as follows. If the electric
field in a uniformly charged (centric) hollow sphere vanishes within
the enclosed space then the internal electric field of a uniformly
charged (single) sphere must be linear dependent on the position.
Indeed if the latter field (say of the outer sphere) is denoted by
$\overline{E}(\overline{r})$, with the origin taken at the center,
then in the enclosed space of a hollow sphere the field equals $\overline{E}(\overline{r})-s\overline{E}(\overline{r}/s)$,
where \emph{s} is the scale factor scaling the outer sphere to the
inner one. If this field vanishes then $\overline{E}(\overline{r})$
must be linear in $\overline{r}$.

Next we transform a spherical shell into a more general ellipsoidal
shell by directional scaling. Taking into account the quadratic dependence
of the electric field on distance it is easy to show that also in
the cavity of a uniformly charged hollow ellipsoid%
\footnote{By a hollow ellipsoid we mean that inner and outer bounding surfaces
are similar ellipsoids, which should not be confused with the case
that the ellipsoids are confocal.%
} the field vanishes \cite{thomson519}. Therefore applying the same
scaling argument as above, the internal field of a uniformly charged
ellipsoid is also linear in the position (referred to the center)
and \eqref{eq:2} can then be written for an arbitrary ellipsoid as:

\begin{equation}
\left\langle \overline{E}_{q}\right\rangle _{\Omega}=\overline{E}_{\rho}(\mathrm{P})=-\frac{q}{\varepsilon_{0}|\Omega|}\overline{\overline{L}}_{i}\cdot\overline{\mathrm{QP}}=-\overline{\overline{L}}_{i}(\overline{\overline{A}})\cdot\frac{\overline{p}_{q}}{\varepsilon_{0}|\Omega|}\qquad\mathrm{P}\in\Omega\label{eq:10}\end{equation}

Where $\overline{\overline{L}}_{i}$ depends on the shape of the ellipsoid,
defined by $\overline{\overline{A}}:\overline{r}\overline{r}=1$,
$\overline{\overline{A}}$ being a symmetric positive definite tensor,
and the index $i$ refers to {}``isotropic'' since we are still
considering an isotropic background. Using \eqref{eq:seven} it follows
that:

\begin{equation}
\left\langle \overline{E}_{\overline{p}}\right\rangle _{\Omega}=\overline{E}_{\overline{P}}(\mathrm{P})=-\overline{\overline{L}}_{i}(\overline{\overline{A}})\cdot\frac{\overline{P}}{\varepsilon_{0}}\qquad\mathrm{P}\in\Omega\label{eq:7}\end{equation}

showing that $\overline{\overline{L}}_{i}$ is the usual depolarizing
tensor. Of course one still needs a method for calculating this tensor.
Today the best known approach is probably by the method {}``separation
of variables'' in ellipsoidal coordinates \cite{landau}, but $\overline{\overline{L}}_{i}$
can also be calculated directly \cite{thomson494}. Perhaps the most
elegant expression can be found by solving the problem in Fourier-space
\cite{milton,michel}:\begin{equation}
\overline{\overline{L}}_{i}(\overline{\overline{A}})=\frac{\overline{\overline{A}}}{4\pi}\cdot\int\frac{\overline{n}\overline{n}}{\overline{\overline{A}}:\overline{n}\overline{n}}d\Omega\label{eq:depolformula}\end{equation}

where $\overline{n}$ is a unit vector in the (spatial) frequency
space and the integration is over the surface of the unit sphere.
The expression immediately reveals that the trace of the depolarizing
tensor equals 1.

Now turning again to a hollow ellipsoidal shell it follows from \eqref{eq:7}
that the field is still zero if the inner ellipsoid becomes eccentric,
but still aligned with the similar outer ellipsoid. This means that
if such an eccentric hollow ellipsoid encloses a dipole, the average
of the field over its volume vanishes. In particular the principal
value integral of the dipole field over an ellipsoidal volume thus
vanishes if the exclusion volume is a uniformly scaled version of
the ellipsoid and irrespective of the position of the dipole%
\footnote{In an appendix of \cite{degroot} this is proved directly and from
this the constancy of $\overline{\overline{L}}_{i}$ is derived.%
}. The corresponding singularity of the dipole field is then $-\overline{\overline{L}}\cdot\frac{\overline{p}}{\varepsilon_{0}}\delta(\overline{r})$
and more in general \eqref{eq:dipole} is written as:\begin{equation}
\varepsilon_{0}\overline{E}_{\overline{p}}=\frac{\overline{p}}{4\pi}\cdot\frac{3\overline{r}\overline{r}-r^{2}\overline{\overline{I}}}{r^{5}}-\overline{\overline{L}}_{i}(\overline{\overline{A}})\cdot\overline{p}\delta(\overline{r})\label{eq:dipoleisotropic}\end{equation}

In this case the dipole is thus modelled more generally as an infinitesimally
small ellipsoid. Obviously there is no unique singularity for a dipole
field and the singularity depends on the symmetry attributed to the
dipole.

The {}``external'' theorem \eqref{eq:1} cannot be extended from
a spherical body to an ellipsoidal body since the field outside a
uniformly charged ellipsoid is not longer equal to that of a monopole,
because the even order multipole moments do not vanish. Consequently
the field outside a uniformly polarized ellipsoid is also not longer
equal to a pure dipole field. However if the ellipsoid is made infinitesimally
small, preserving the shape and keeping the dipole moment constant,
then all higher order moments of the uniform charge distribution will
vanish and the external field of the polarized body once more becomes
a pure dipole field, independent of the shape of the ellipsoid. This
explains why the non-singular part in \eqref{eq:dipoleisotropic}
does not depend on the shape.

\section{\label{sec:Anisotropic-medium.}Anisotropic medium.}

The results can also be extended to a uniform anisotropic background
medium. E.g. consider a uniform dielectric medium with symmetrical
relative dielectric tensor $\overline{\overline{\varepsilon}}$. The
extension is easily handled by introducing a suitable coordinate transformation
which transforms the anisotropic medium into an isotropic one and
then applying the results for the isotropic medium \cite{sihvola}.
If we preserve the electric potential and charge density in corresponding
points then the relevant transformation formulas are given by:\begin{alignat}{1}
\overline{r} & =\overline{\overline{T}}^{1/2}\cdot\overline{r}'\nonumber \\
\overline{E} & =\overline{\overline{T}}^{-1/2}\cdot\overline{E}'\nonumber \\
\overline{D} & =\overline{\overline{T}}^{1/2}\cdot\overline{D}'\nonumber \\
\overline{\overline{\varepsilon}} & =\overline{\overline{T}}^{1/2}\cdot\overline{\overline{\varepsilon}}'\cdot\overline{\overline{T}}^{1/2}\nonumber \\
q & =|\overline{\overline{T}}|^{1/2}q'\nonumber \\
\overline{p} & =|\overline{\overline{T}}|^{1/2}\overline{\overline{T}}^{1/2}\cdot\overline{p}'\nonumber \\
\overline{\overline{A}} & =\overline{\overline{T}}^{-1/2}\cdot\overline{\overline{A}}'\cdot\overline{\overline{T}}^{-1/2}\nonumber \\
\overline{\overline{L}} & =\overline{\overline{T}}^{1/2}\cdot\overline{\overline{L}}'\cdot\overline{\overline{T}}^{-1/2}\label{eq:transformations}\end{alignat}

where $|\overline{\overline{T}}|$ is the determinant of $\overline{\overline{T}}$.
With $\overline{\overline{T}}=\overline{\overline{\varepsilon}}$
the transformed dielectric tensor becomes isotropic ($\overline{\overline{\varepsilon}}'=\overline{\overline{I}}$).
Consider now as volume $\Omega$ an ellipsoid isomorphous with the
{}``index ellipsoid'' of the medium $\overline{\overline{\varepsilon}}^{-1}:\overline{r}\overline{r}=1$,
then $\Omega$ is transformed into a sphere $\Omega'$ in an isotropic
medium, to which \eqref{eq:1} can be applied. Transforming back we
find that \eqref{eq:1} remains valid for an anisotropic medium when
applied to a volume isomorphous with the {}``index ellipsoid''.
The depolarizing tensor $\overline{\overline{L}}(\overline{\overline{\varepsilon}},\overline{\overline{A}})$
depends on the anisotropy of the medium and on the shape of the ellipsoidal
volume and follows directly from the last two transformation formula's
in \eqref{eq:transformations}:\begin{equation}
\overline{\overline{L}}(\overline{\overline{\varepsilon}},\overline{\overline{A}})=\overline{\overline{\varepsilon}}^{1/2}\cdot\overline{\overline{L}}_{i}(\overline{\overline{\varepsilon}}^{1/2}\cdot\overline{\overline{A}}\cdot\overline{\overline{\varepsilon}}^{1/2})\cdot\overline{\overline{\varepsilon}}^{-1/2}\label{eq:anisotropic1}\end{equation}

and \eqref{eq:10} and \eqref{eq:7} can thus be extended to:\begin{alignat}{1}
\varepsilon_{0}\overline{\overline{\varepsilon}}\cdot\left\langle \overline{E}_{q}\right\rangle _{\Omega} & =-\overline{\overline{L}}(\overline{\overline{\varepsilon}},\overline{\overline{A}})\cdot\frac{\overline{p}_{q}}{|\Omega|}\qquad\mathrm{P}\in\Omega\label{eq:textended}\\
\varepsilon_{0}\overline{\overline{\varepsilon}}\cdot\left\langle \overline{E}_{\overline{p}}\right\rangle _{\Omega} & =-\overline{\overline{L}}(\overline{\overline{\varepsilon}},\overline{\overline{A}})\cdot\frac{\overline{p}}{|\Omega|}\qquad\mathrm{P}\in\Omega\nonumber \end{alignat}

showing that the second theorem actually holds for the electric displacement
instead of for the electric field. From \eqref{eq:anisotropic1} it
also follows that the trace of $\overline{\overline{L}}(\overline{\overline{\varepsilon}},\overline{\overline{A}})$
is still 1.

As a special case we consider now a sphere ($\overline{\overline{A}}=\overline{\overline{I}}$)
and we then obtain in particular $\overline{\overline{L}}(\overline{\overline{\varepsilon}},\overline{\overline{I}})=\overline{\overline{\varepsilon}}^{1/2}\cdot\overline{\overline{L}}_{i}(\overline{\overline{\varepsilon}})\cdot\overline{\overline{\varepsilon}}^{-1/2}=\overline{\overline{L}}_{i}(\overline{\overline{\varepsilon}})$
as given by \eqref{eq:depolformula} and where in the last step we
used the property that $\overline{\overline{L}}_{i}(\overline{\overline{\varepsilon}})$
and $\overline{\overline{\varepsilon}}$ share the same principal
axes. It follows that $\overline{\overline{\varepsilon}}$ and $\overline{\overline{A}}$
separately have the same effect on the depolarizing tensor. We consider
then an alternative transformation which transforms the ellipsoid
into a sphere ($\overline{\overline{T}}=\overline{\overline{A}}^{-1}$
and $\overline{\overline{A}}'=\overline{\overline{I}}$) and obtain
in this way an alternative expression for the depolarizing tensor:\begin{equation}
\overline{\overline{L}}(\overline{\overline{\varepsilon}},\overline{\overline{A}})=\overline{\overline{A}}^{-1/2}\cdot\overline{\overline{L}}_{i}(\overline{\overline{A}}^{1/2}\cdot\overline{\overline{\varepsilon}}\cdot\overline{\overline{A}}^{1/2})\cdot\overline{\overline{A}}^{1/2}\label{eq:anisotropic2}\end{equation}

The expression \eqref{eq:anisotropic2} with \eqref{eq:depolformula}
has been further extended to a bianisotropic medium \cite{weiglhofer1999}.

Comparing \eqref{eq:anisotropic1} and \eqref{eq:anisotropic2} it
follows that: \begin{alignat}{1}
\overline{\overline{L}}(\overline{\overline{\varepsilon}},\overline{\overline{A}})\cdot\overline{\overline{\varepsilon}} & =\overline{\overline{\varepsilon}}\cdot\overline{\overline{L}}(\overline{\overline{A}},\overline{\overline{\varepsilon}})\nonumber \\
\overline{\overline{A}}\cdot\overline{\overline{L}}(\overline{\overline{\varepsilon}},\overline{\overline{A}}) & =\overline{\overline{L}}(\overline{\overline{A}},\overline{\overline{\varepsilon}})\cdot\overline{\overline{A}}\end{alignat}
showing the more general symmetry between anisotropy ($\overline{\overline{\varepsilon}}$)
and shape ($\overline{\overline{A}}$). The non-singular part of the
dipole field can also be obtained with the first coordinate transformation
resulting into the following dipole field:\begin{equation}
\varepsilon_{0}\overline{\overline{\varepsilon}}\cdot\overline{E}_{\overline{p}}=\frac{\overline{p}}{4\pi|\overline{\overline{\varepsilon}}|^{1/2}}\cdot\left[\frac{3\overline{\overline{\varepsilon}}^{-1}\cdot\overline{r}\overline{r}}{\left(\overline{\overline{\varepsilon}}^{-1}:\overline{r}\overline{r}\right)^{5/2}}-\frac{\overline{\overline{I}}}{\left(\overline{\overline{\varepsilon}}^{-1}:\overline{r}\overline{r}\right)^{3/2}}\right]-\overline{\overline{L}}(\overline{\overline{\varepsilon}},\overline{\overline{A}})\cdot\overline{p}\delta(\overline{r})\label{eq:final}\end{equation}

which reduces to \eqref{eq:dipoleisotropic} for $\overline{\overline{\varepsilon}}=\overline{\overline{I}}$.
Usually this is called a Green's dyadic and the field of an arbitrary
polarization density $\overline{P}(\overline{r})$ is then obtained
by a convolution with this dyadic, with the last term resulting into
$-\overline{\overline{L}}(\overline{\overline{\varepsilon}},\overline{\overline{A}})\cdot\overline{P}$
which is only present if the observation point is in the source region.
The link between this contribution and the depolarizing tensor has
been noted by Weiglhofer \cite{weiglhofer2000}.

\end{document}